\documentclass[prl,aps,twocolumn,amsmath,amssymb,superscriptaddress]{revtex4-1}
\usepackage[dvipsnames]{xcolor}
\usepackage{amssymb}
\usepackage{esint}

\usepackage{amstext}
\usepackage{graphicx,epstopdf}
\usepackage{color}
\usepackage{amsmath}
\usepackage{color}
\usepackage{enumitem}
\usepackage{soul}
\usepackage{makecell}

\makeatletter

\definecolor{amber}{rgb}{1.0, 0.75, 0.0}

\newcommand{\abs}[1]{\left< #1\right>}

\begin{document}

	\title{
		Diffusion through a network of compartments separated by partially-transmitting boundaries
		}
	
	\author{G. Mu\~noz-Gil}
	\affiliation{ICFO -- Institut de Ci\`encies Fot\`oniques, The Barcelona Institute of Science and Technology, 08860 Castelldefels (Barcelona), Spain}
	\author{M.A. Garcia-March} 
	\affiliation{ICFO -- Institut de Ci\`encies Fot\`oniques, The Barcelona Institute of Science and Technology, 08860 Castelldefels (Barcelona), Spain}
	\author{C. Manzo}	 
	\affiliation{Universitat de Vic -- Universitat Central de Catalunya (UVic-UCC), C. de la Laura,13, 08500 Vic, Spain}
	\author{A. Celi}
	\affiliation{Center for Quantum Physics, University of Innsbruck, and Institute for Quantum Optics and Quantum Information, Austrian Academy of Sciences, Innsbruck, Austria} 
	\affiliation{ICFO -- Institut de Ci\`encies Fot\`oniques, The Barcelona Institute of Science and Technology, 08860 Castelldefels (Barcelona), Spain}
	\author{M. Lewenstein}
        \affiliation{ICFO -- Institut de Ci\`encies Fot\`oniques, The Barcelona Institute of Science and Technology, 08860 Castelldefels (Barcelona), Spain}
        \affiliation{ICREA, Lluis Companys 23, E-08010 Barcelona, Spain}

	\begin{abstract}
	We study the random walk of a particle in a compartmentalized environment, as realized in biological samples or solid state compounds. Each compartment is characterized by its length $L$ and the boundaries transmittance $T$. We identify two relevant spatio-temporal scales that provide alternative descriptions of the dynamics: i) the microscale, in which the particle position is monitored at constant time intervals; and ii) the mesoscale, in which it is monitored only when the particle crosses a boundary between compartments. Both descriptions provide --by construction-- the same long time behavior.  The analytical description obtained at the proposed mesoscale allows for a complete characterization of the complex movement at the microscale, thus representing a fruitful approach for this kind of systems. We show that the presence of disorder in the transmittances is a necessary condition to induce anomalous diffusion, whereas the spatial heterogeneity reduces the degree of subdiffusion and, in some cases, can even compensate for the disorder induced by the stochastic transmittance. 
	\end{abstract}
	\maketitle

\paragraph*{Introduction} 


The characterization of the diffusive behavior in complex environments is crucial in many fields, ranging from biology~\cite{2018Tan} to geology~\cite{2006Berkowitz}. Recently, it has been shown that a large number of systems display anomalous diffusion associated to spatial and/or energetic disorder of the environment. Often, the motion of particles in such systems has been shown to be subdiffusive, i.e.  $\left<x^2(t)\right>\sim t^\sigma$ with anomalous exponent $0<\sigma<1$. The characterization of this movement provides important information on the disorder of the media and on the laws governing the system~\cite{2014Metzler}.  The advances in this field have been mainly driven by developments in fluorescence microscopy, which enable us to record movies of single particles diffusing in living matter, with a spatial precision of a few nanometers at the millisecond time scale~\cite{Manzo2015ROPP}.

\begin{figure}
	\includegraphics[width=0.8\columnwidth]{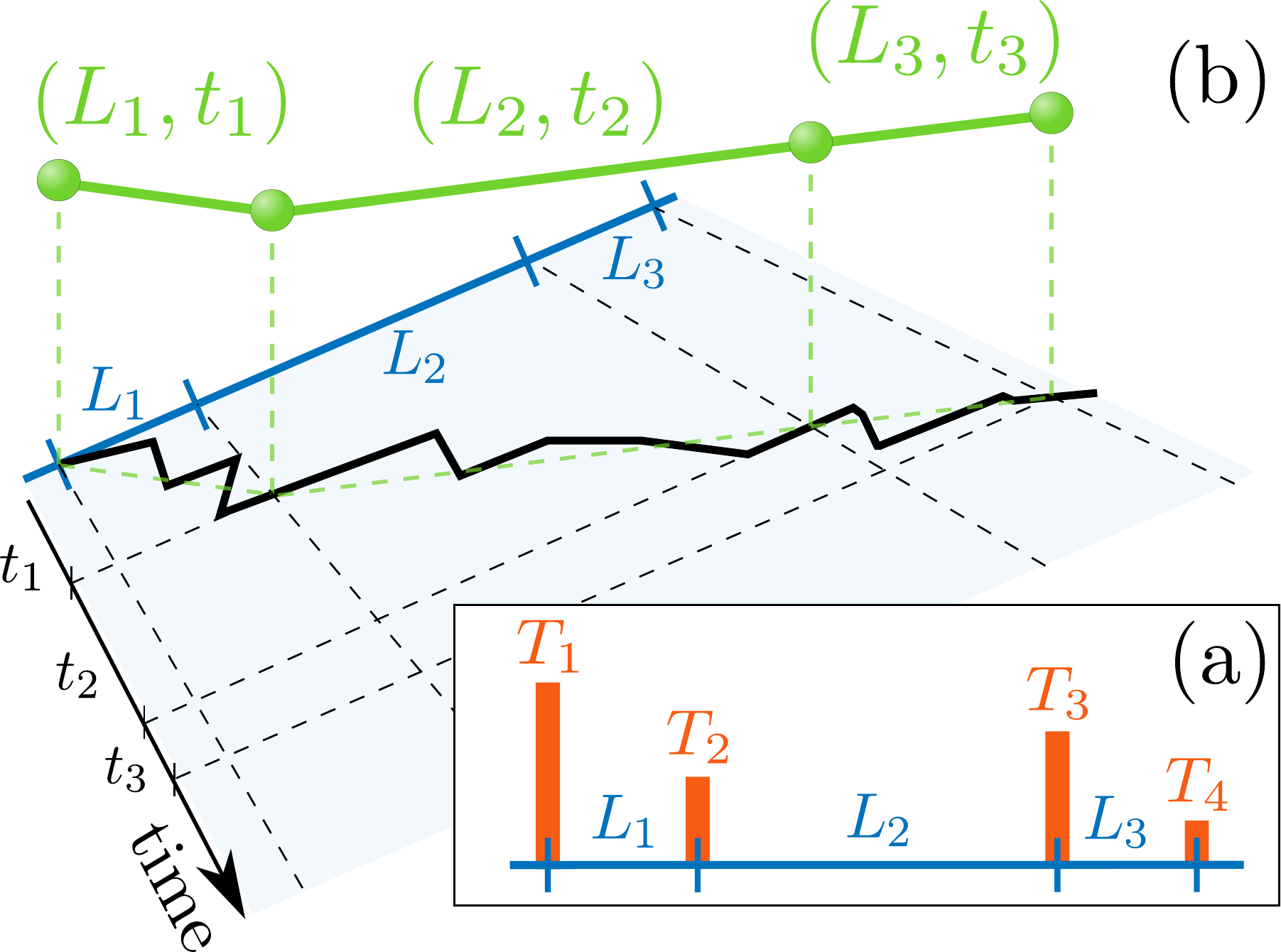}
	\caption{{\it Schematic of the system}. (a) Example of a one dimensional compartmentalized environment, with compartment size $L$ and boundary transmittance $T$. Higher boundaries represent lower transmittance. For simplicity we plot the meshwork as formed by the segments of a line. (b) Motion of the particle in such environment. The dark line represents the microscale description of the motion and the green one indicates   its mesoscale description, in this case a L\'evy walk with steps given by their length and flight time $(L, t)$.   
		\label{fig:fig1}  }
\end{figure}

The presence of barriers that prevent the particles to freely diffuse in the environments is a general mechanism used to explain subdiffusion~\cite{Haus1987}. Indeed, there exists a plethora of works treating the effect of these barriers in various forms, from local maxima in potential landscapes~\cite{Bernasconi1979} to thin slices of poorly diffusive materials~\cite{Novikov2011}. Recent experimental observations in cellular biology have shown that the actin cytoskeleton acts as a compartmentalization scaffold for proteins diffusing in the plasma membrane~\cite{Sadegh2017,deWit2018}, hence stressing the importance of studying the motion in such environments. Moreover, the evidence of the occurrence of ergodic and nonergodic processes in the diffusion of biomolecules~\cite{Weigel2011} has triggered the description of models in which geometric and energetic disorders coexist~\cite{Meroz2010}.

In this Letter, we study a general barrier model, where a particle performs an unbiased random walk through a complex environment made by a mesh of compartments separated by barriers with random transmittance. A schematic of the system is shown in Fig.~\ref{fig:fig1}.  We show that even though the particle performs a Brownian motion within each compartment, the stochasticity of the barrier's transmittance induces anomalous diffusion for the overall movement. We also explore the effect of the stochasticity in the length of the compartments, showing that it generally increase the anomalous exponent, up to restoring normal diffusion. .

In order to study the behavior of the particle, we propose a coarse-graining approach transforming the rather complex walk of the particle (mainly due to the interaction with the boundaries) into either a continuous time random walk or a L\'evy walk,  which provides a simpler description of the process from the mathematical point of view.  In the most general description of our system, we show how the walk of the particle can be mapped into a L\'evy walk with rests, where flight times depend on the step size. 
In our system, the steps and rests are not alternate but have complementary probabilities at each event. We show how the existing theory for a L\'evy walk with rests can be extended to study such kind of walk. We determine the relationship between the stochasticity of the environment and the anomalous diffusion of the particle by solving different configurations of our system, characterized by fixed or random compartment sizes and boundary transmittances.

\paragraph*{The Model} The motion takes place on an environment characterized by a set of compartments with size $\left\{L_i\right\}_{i=1}^N$, with $N\gg1$ and $L_i\in\left[1,\infty \right)$. We treat the size of the compartments as a stochastic variable, following the probability distribution function (PDF) $g(L)$. The compartments form a meshwork with unbounded connectivity, which we assume to be always sufficiently large such to make very unlikely that the particle returns to the same compartment after leaving it. The boundary between the compartments is partially reflective, i.e. a particle reaching a boundary has a finite probability $T$ of moving through the boundary to the next compartment and a complementary probability $R=1-T$ of being reflected. The transmittance of each segment $\left\{T_{i}\right\}_{i=1}^N$, $T\in\left(0,1 \right]$ is a random variable drawn from the PDF $q(T)$.

For the sake of simplicity, we  focus on the case where the compartments consist in one-dimensional segments [see Fig.~\ref{fig:fig1}(a)]. The extension of this theory to two- or three-dimensional supports, like circles or spheres, is conceptually straightforward but more elaborated and geometry-dependent, since it requires the determination of the stochastic time that the particle spends in each support. The particle performs an unbiased, discrete, random walk through the environment, temporarily confined between the boundaries until it is transmitted to the next compartment. 

\paragraph*{Methods} The motion of particles in disordered media has been thoroughly studied in the past~\cite{Bouchaud1990}. The usual approach is to explicitly solve the diffusion equation for the system under study. For instance, such direct approach has been recently applied to subdiffusive particles through the barrier separating two liquids~\cite{Kosztolowicz2017}. However, when considering systems like the one presented above, where both the boundary transmittance and compartment length are stochastic variables, the direct approach is complicated and does not lead to exact analytical results. Therefore, we use an alternative method to solve the motion of the particle through such a system. First, we distinguish between a microscale description, in which the position of the particle is monitored at constant times $<<L^2/D$ with $D$ being the diffusivity, and a mesoscale description, in which the position is sampled at times subordinated to the exit from a compartment. We note here that, by definition, the asymptotic behavior of the motion of the particle coincides on both scales. Therefore, studying the movement at the mesoscale provides a correct description of the movement at long times. 

In the mesoscale description, the microscopic walk of the particle (represented by the black line in the same figure) is reduced  to a collection of lengths ($L_i$) and times ($t_i$) traveled to exit the compartments, as shown by the green line of Fig.~\ref{fig:fig1}(b). As a matter of fact, the length traveled by the particle in each step corresponds to the size of the compartment itself. The {\it flight time} $t_i$ is the stochastic time the particle spent bouncing between the boundaries before being transmitted to next compartment. In our case, this time is related to the transmittance $T$ and length $L$ of the compartment through the conditional probability $\phi(t|T,L)$. One can then write the joint probability for the particle to be in a compartment of length $L$ and boundary transmittance $T$ at  time $t$ as
\begin{equation}
\label{eq:joint}
\psi(t,L,T)=\phi(t|T,L)g(L)q(T).
\end{equation}
Once inside a compartment, the particle has two options: leaving through the same boundary through which it entered, or through the opposite one. Since our approach monitors the particle only when exiting a boundary, in the latter case, the particle has traveled a distance equal to the size of the compartment. However, in the former, the particle is not effectively moving, since it occupies the same position when entering and exiting the compartment. This translates into a rest with duration equal to the time taken to exit the compartment. Therefore, after entering each compartment, the particle has a probability of resting $\varphi_r(L,T)$ and the complementary probability of walking $\varphi_w(L,T)=1-\varphi_r(L,T)$.

Through this coarse-graining approach, we convert the microscale walk into a L\'evy walk with rests, with flight times depending on the jump length~\cite{Zaburdaev2006}. Previous works have extensively studied  such kind of walks, both with alternating walks and rests~\cite{Sokolov2011} or with an equal probability of resting and walking~\cite{Zaburdaev2015}. However, our system shows a substantial difference, since it displays different probabilities of resting or walking, $\varphi_r+\varphi_w=1$,  that can be used to calculate the PDFs of walk [$\psi_w(t)$] and rest times [$\psi_r(t)$] as
\begin{equation}
\label{eq:walkresttimes}
\psi_{w(r)}(t) \!=\!\!\!\int^{\infty}_{1}dL \int^{1}_{0}\varphi_{w(r)}(L,T) \psi(t,L,T) dT,
\end{equation}
and, in the spirit of \cite{Zaburdaev2015}, to derive the density of particles at position $x$ and time $t$ in the Fourier--Laplace space  
\begin{equation}
P_{\Sigma}(k,s)=\int_0^1 P_{\Sigma,T}(k,s,T) dT,
\end{equation}
where
\begin{align}
\label{eq:total_density}
& P_{\Sigma,T}(k,s,T) =\\
& \frac{\Psi_r(s)P_0(k)+\left\{\varphi_w(x,T)\Psi(x,s,T)\right\}_k\psi_r(s)P_0(k)}{1-\{\varphi_r(x,T))\psi(x,s,T)\}_k\psi_r(s)}.\nonumber
\end{align}
Here, $P_0(x)$ corresponds to the initial distribution of particles, $\Psi(t) = \int_t^\infty\psi(t')dt'$ to the survival probability, 
i.e. the probability of not jumping until time $t$, $\Psi(x,t,T)=\int_t^\infty\psi(x,t',T)dt'$ to the PDF of the displacement of the walker during the last uncompleted step, 
and $\{f(x)\}_k$ to the Fourier transform of $f(x)$. 
For constant step/rest probabilities, e.g. $\varphi_w=\varphi_r=1/2$, Eq.~\eqref{eq:total_density} leads to the known result for the L\'evy walk with rests~\cite{Zaburdaev2015}. 

However, when the previous condition is not fulfilled, solving Eq.~\eqref{eq:total_density} requires the calculation of $\varphi_w(L,T)$.  A case in which $\varphi_w(L,T)$ is easily solvable is when the boundaries are completely transmitting, i.e. $q(T)=\delta(T-1)$. In that case, one finds
\begin{equation}
\label{eq:prob_walk_T1}
\varphi_w(L,T=1)=\varphi_w(L)=1-\frac{L}{L+1}\sim L^{-1}.
\end{equation}
For $T\neq1$, obtaining an analytical expression for $\varphi_w(L,T)$ is a challenging task~\cite{Lehner1963}. A trick commonly used to avoid this difficulty consists in considering an {\it annealed} system~\cite{Bouchaud1990}, i.e. assuming that each time the particle exits a compartment, it reappears at the center of the next one. In this case, the particle will always travel a distance $L_i/2$ to escape the $i$th-segment, independently on the exit side, hence eliminating the presence of rests. In this case $\varphi_w(L,T)=1 \ \forall L,T$ and the motion of the particle is then a L\'evy walk with flying times depending on the jump length~\cite{Zaburdaev2006}. This is also analogous to the case in which, once the particle enters a compartment, it cannot cross again the same edge it entered from and thus will always travel a distance $L_i$. For this reason, in the following we will refer to this approximation as the \textit{osmotic}  approach, in contrast with the general case that we name \textit{non-osmotic}. 

From now on we will focus on the osmotic approach, which allows for a thorough theoretical description in the different configurations considered. In the osmotic approach, Eq.~\eqref{eq:total_density} takes the much simpler form
\begin{equation}
\label{eq:total_density_OA}
P^{\mathrm{(OA)}}_\Sigma = \frac{\Psi(k,s)}{1-\psi(k,s)},
\end{equation}
where $\psi(k,s) \!=\!\int_0^1 \!\!\psi(k,s,T)dT$.

To characterize the motion of the particle, we will use the mean squared displacement (MSD), defined as $\abs{x^2(t)}=-P''(k,s)|_{k=0}$, which can be rewritten as~\cite{Massignan2014}
\begin{equation}
\abs{x^2(s)} = \int_0^1 dT \left[\frac{-\psi''(k,s)|_ {k=0}}{s[1-\psi_w(s)]}
+\frac{-\Psi''(k,s)|_ {k=0}}{1-\psi_w(s)}\right].
\label{eq:msd_patch}
\end{equation}

As we will show later through numerical simulations of the microscopic walk, in spite of the simpler description, the osmotic approach displays  the same long time behavior as the non-osmotic one.

\paragraph*{Results} In the following, we will use the method described above to solve the motion of the particle in different configurations of the system. We will first consider the case in which each boundary has a different transmittance, drawn stochastically from the PDF $q(T)$, but all the compartments have equal length. We will then briefly comment about the case in which the stochasticity is only present in the compartment length. Last, we will consider the case where both the length and boundary transmittance are random variables. For each case, we will give the analytical solutions of the mesoscopic walk and compare it to numerical simulations of the microscopic description.

The form of the conditional probability of the exit time given a compartment of size $L$ and transmittance $T$ is common to all the cases.  A reasonable assumption based on the Brownian motion is that, independently on the expression of this conditional probability, it should give an average time for exiting a compartment $ \langle t\rangle$ which depends on the length as $L^2$. We can further assume that the dependence on $T$ is such that $ \langle t\rangle\propto(L/T)^2$. We checked that this behavior is consistent with the numerical results for a collection of $T$ and $L$, finding that the average exit time follows an exponential behavior, $\propto\exp{-t L^2/T^2}$, for large $L$  and small $T$. For all cases numerically considered, even when the distribution did not match an exponential behavior, we found a quadratic dependence on $L/T$. Therefore, we assume the simplest distribution which produces the expected behavior of the average exit time, which is
\begin{equation}
\phi(t|T,L)\sim\delta(t-(L/T)^2).
\end{equation}
This form of the conditional time also has the advantage of simplifying the analytical expressions and,  as we discuss below, allows us to correctly model the microscopic motion in all the cases considered. The analytical calculation of this conditional probability falls beyond the scope of this Letter. We note that previous works have focused in the investigation on the exit time in similar structures~\cite{Khantha1983, Dybiec2006}, but do not provide a useful expression for our particular system nor a practical way to derive it.  

We will now consider the case in which the boundaries have all the same transmittance, i.e. $q(T)=\delta(T-\bar{T})$, with $\bar{T}\in(0, 1]$. We will consider that each compartment has a different length, retrieved from the PDF
\begin{equation}
\label{eq:probx}
g(L)=\beta L^{-1-\beta}.
\end{equation}
Our first step is to calculate the distribution of flight times, which is done similarly to Eq.~\eqref{eq:waiting_CTRW} by convolving Eq.~\eqref{eq:joint} over all possible values of $L$
\begin{equation}
\label{eq:times_CTRL}
\psi_L(t)=\int_1^\infty \phi(t|L,\bar{T}) g(L) dL \propto  t^{-1-\beta/2}.
\end{equation}
Using this result and Eqs.~\eqref{eq:joint} and~\eqref{eq:msd_patch}, we find that $\abs{x^2(t)}\sim t \ \forall \ T$, i.e. the particle performs normal diffusion. Therefore, the stochasticity of the compartment length does not imply any effect on the MSD and, indeed, similar results are obtained when using regular compartment size.  We would like to emphasize that this result holds for any finite $T$ different from zero. In fact, as shown in~\cite{Lapeyre2015}, for $T=0$ subdiffusion occurs. In the presence of transmitting boundary, there is no mechanism that confines the particle for pathologically long times, so particles diffuse normally in the asymptotic limit. 

\begin{figure}
	\begin{center}
		\includegraphics[width=0.95\columnwidth]{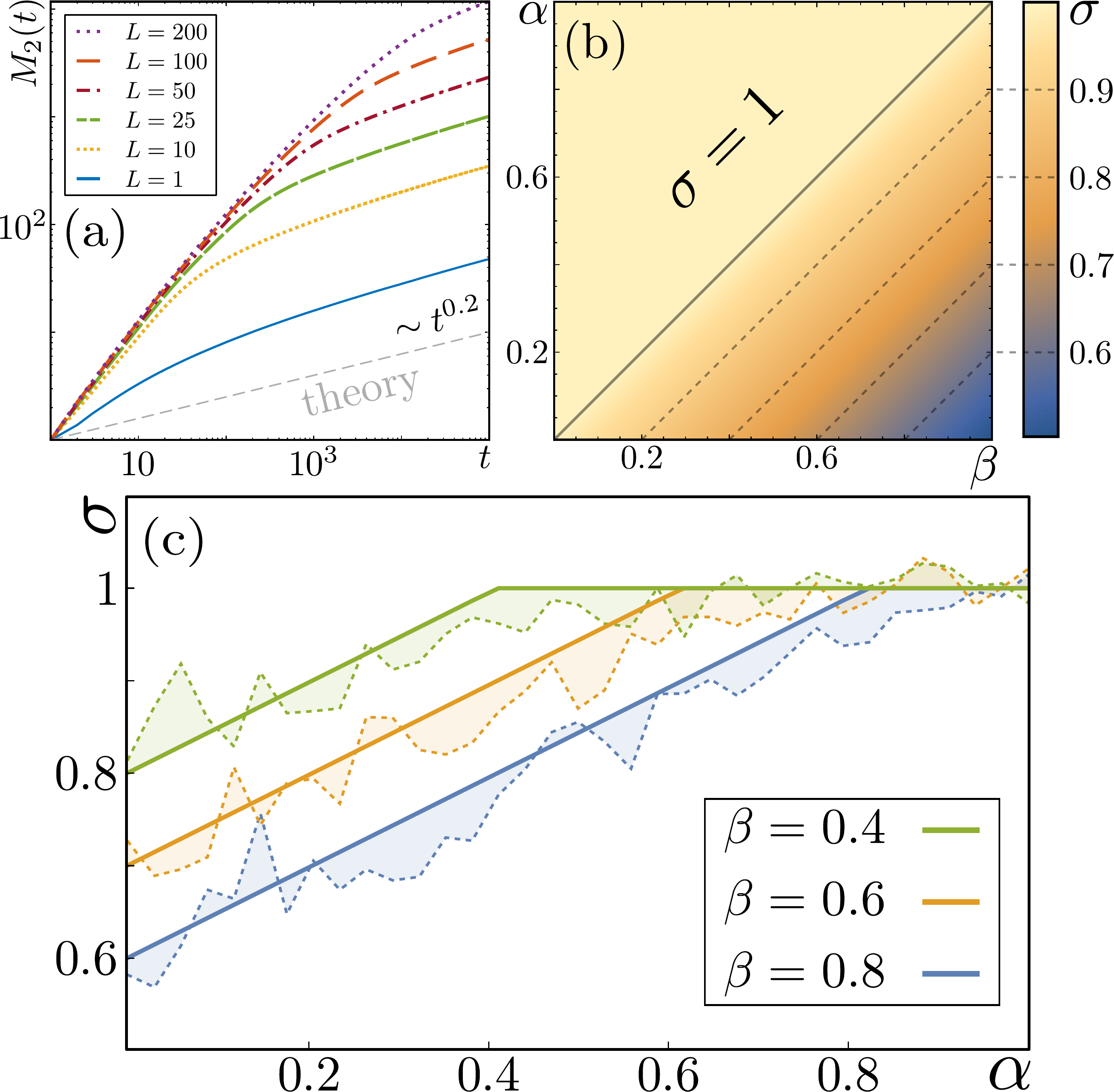}
		\caption{(a) MSD of a particle moving in an system of compartments of equal length and boundary transmittances distributed following \eqref{eq:prob_T}, with $\alpha = 0.2$. All curves are calculated for the microscale and tend to the predicted subdiffusive motion given by \eqref{eq:msdfixed}. A larger $L$ leads to a larger time for the onset of subdiffusion to occur. The dashed bottom line corresponds to the mesoscale and coincides with the theoretical prediction. (b) Value of the exponent of the MSD in a system with stochastic compartment sizes and boundary transmittance, given by \eqref{eq:MSD_RTRL}. (c) Comparison between the predicted results of the previous case and numerical simulations of the microscopic walk (dashed lines).
			\label{fig2}}
	\end{center}
\end{figure}

A very different result arises when considering disordered boundary transmittances $\left\{T_{i}\right\}_{i=1}^N$ distributed according to a power law PDF
\begin{equation}
\label{eq:prob_T}
q(T)=\alpha\left(\frac{1}{T}\right)^{1-\alpha}.
\end{equation}
We first analyze the case in which the compartments have all the same size, i.e. the lengths $\left\{L_i\right\}_{i=1}^N$ are distributed according to the PDF $g(L)=\delta(L-\bar{L})$, where $\bar{L}\in[1,\infty)$. We refer to this system as the spatially ordered case. In the osmotic approach, the walk consists on a collection of steps of size $\bar{L}$ with flight times drawn from the PDF
\begin{equation}
\label{eq:waiting_CTRW}
\psi(t)=\int_0^1 \phi(t|\bar{L},T) q(T) dT \propto  t^{-1-\alpha/2}.
\end{equation}
As all the steps have equal length, the walk reduces to a continuous time random walk with waiting time PDF given by \eqref{eq:waiting_CTRW}. Thus, in the spatially ordered case the MSD is given by~\cite{Charalambous2017}  
\begin{equation}
\label{eq:msdfixed}
\abs{x^2(t)}^{(\mathrm{SO})}\xrightarrow{t\rightarrow\infty} t^{\alpha/2},
\end{equation}
showing that the particle undergoes subdiffusive motion for $0<\alpha<1$. In Fig.~\ref{fig2} (a) we show the numerical results corresponding to MSD calculated for a single value of $\alpha=0.2$ and different values of $\bar{L}$ by using the microscale description for the spatially ordered case. The plot shows that the motion is initially Brownian and become subdiffusive at longer times. The time at which the onset of subdiffusion occurs increases as $\bar{L}$ grows, corresponding to the time needed to reach the boundary, of a compartment. The asymptotic value of the MSD for any $\bar{L}$ is given by \eqref{eq:msdfixed}. This is a first indication that anomalous diffusion can only be obtained by considering stochastic boundary transmittance with a heavy-tail PDF. In the spatially ordered case, the distribution of transmittances of the media can be directly inferred from the asymptotic behavior of the MSD of the particle. 

We will now consider the case where both compartment length and boundary transmittance are stochastic variables. As stated before, this situation can be modeled at the mesoscale as a L\'evy walk with flight times depending on the step size. We consider that the transmittances are distributed according to Eq.~\eqref{eq:prob_T} and the compartment lengths as described by Eq.~\eqref{eq:probx}. Following the method used to derive  Eq.~\eqref{eq:waiting_CTRW}, we can calculate the PDF of flight times by convolving the conditional probability $\phi(t|T,L)$ with Eqs.~\eqref{eq:prob_T} and~\eqref{eq:probx}, to find
\begin{equation}
\label{eq:times_RLRT}
\psi_f(t)\propto t^{-1-\gamma} ,\ \ \ \mbox{with } \gamma = 
\begin{cases}
\alpha & \mbox{if } \beta>\alpha, \\
\beta & \mbox{if } \beta<\alpha.
\end{cases} 
\end{equation}
By using the previous result and Eq.~ \eqref{eq:joint} we can determine the MSD through its Laplace transform as in Eq.~\eqref{eq:msd_patch}.  In the time domain we find   
\begin{equation}
\label{eq:MSD_RTRL}
\abs{x^2(t)}^{(\mathrm{SD})}\xrightarrow{t\rightarrow\infty} t^{\frac{1}{2}(2-\beta+\gamma)}.
\end{equation}
The values of the MSD exponent $\sigma =\frac{1}{2}( 2-\beta+\gamma)$ obtained for different values of $\alpha$ and $\beta$ are shown in Fig.~\ref{fig2}(b). In Fig.~\ref{fig2}(c) we further show the values of the MSD exponent calculated from numerical simulations for the microscale description of the walk (dashed lines) and the theoretical value given by Eq.~\eqref{eq:MSD_RTRL}. The numerical calculation and the theoretical prediction show a good agreement. It can be noticed that, when $\alpha>\beta$ (and thus $\gamma=\beta$), the particle movement is normally diffusive [see  Eq.\eqref{eq:MSD_RTRL} and Fig.~\ref{fig2}(b) and (c)]. Therefore, the stochasticity in the length of the segments  is capable of compensating for the disorder that would be induced by the stochasticity in transmittance, that would generate a subdiffusive motion with anomalous exponent $\sigma=\alpha/2$ in the case the segments lengths were regular. In addition,  for $\beta>\alpha$, the motion is subdiffusive, but with a higher anomalous exponent as compared to the case in which the lengths were regular. Therefore in this case the two disorders compete, producing a weaker subdiffusion.

\paragraph*{Conclusions} In this Letter we show that mesoscopic approaches based on Levy walk or continuous time random walk often used to describe anomalous diffusion, can be linked to realistic microscopic behavior such as the diffusion in a compartmentalized environment with random sizes and/or transmittances. We show that a heavy-tailed distribution of transmittances is necessary to obtain a subdiffusive motion. We further demonstrate that, while geometric disorder cannot generate subdiffusion by itself, it can affect the one generated by the heterogeneity in the boundary transmittance and increase the value of the anomalous exponent toward one and thus inducing a smaller degree of subdiffusion. The model presented in this Letter might be an useful framework to interpret the motion of a variety of transmembrane proteins diffusing at the cell surface for which actin cytoskeleton filaments act as semipermeable barriers.

\acknowledgments
We acknowledge Oriol Bard\'es for the initial numerical exploration of the problem and John Lapeyre and Vasily Zaburdaev for inspiring and useful discussions.  This work has been funded by the Spanish Ministry MINECO (National Plan15 Grant: FISICATEAMO No. FIS2016-79508-P, SEVERO OCHOA No. SEV-2015-0522, FPI), European Social Fund, Fundaci\'o  Cellex, Generalitat de Catalunya (AGAUR Grant No. 2017 SGR 1341 and CERCA/Program), ERC AdG OSYRIS, EU FETPRO QUIC, and the National Science Centre, Poland-Symfonia Grant No. 2016/20/W/ST4/00314. C.M. acknowledges funding from the Spanish Ministry of Economy and Competitiveness and the European Social Fund through the Ram\'{o}n y Cajal program 2015 (RYC-2015-17896) and the BFU2017-85693-R, and from the Generalitat de Catalunya (AGAUR Grant No. 2017SGR940). AC acknowledges financial support from the ERC Synergy Grant UQUAM and the SFB FoQuS (FWF Project No. F4016-N23).

\end{document}